\documentclass[a4paper,12pt]{article}

\usepackage[utf8x]{inputenc}
\usepackage[T1]{fontenc}
\usepackage{times}
\usepackage[colorlinks=true,linkcolor=black,citecolor=black,urlcolor=blue]{hyperref}
\usepackage{geometry}
\usepackage{titlesec}
\titleformat{\section}
{\bfseries\uppercase}{\thesection.}{1em}{}
\titleformat{\subsection}
{\bfseries}{\thesection.\thesubsection.}{1em}{}

\usepackage{graphicx} % used to insert the figure
\usepackage{multirow} % used for the table
\usepackage{cite}
\usepackage{breakurl}
\usepackage{indentfirst}
\usepackage{amsmath, amssymb, amsfonts, bm}
\usepackage{txfonts}
\usepackage{enumitem}
\usepackage{color,xcolor}
\usepackage{colortbl}
\usepackage{threeparttable}
\usepackage{subfigure}

\definecolor{myred}{RGB}{228,26,28}
\definecolor{myblue1}{RGB}{166, 206, 227}
\definecolor{myblue2}{RGB}{31,120,180}

\titlespacing*{\subsection}{0pt}{1.5em}{0.2em}

\renewcommand\eqref[1]{Equation~\ref{#1}}

\renewcommand{\thesection}{\arabic{section}}
\renewcommand{\thesubsection}{\arabic{subsection}}

\newlength{\bibitemsep}
\newlength{\bibparskip}
\let\oldthebibliography\thebibliography
\renewcommand\thebibliography[1]{%
  \oldthebibliography{#1}%
  \setlength{\parskip}{\bibitemsep}%
  \setlength{\itemsep}{\bibparskip}%
}
\begin{document}

\begin{center}
	\includegraphics[width=3.50in]{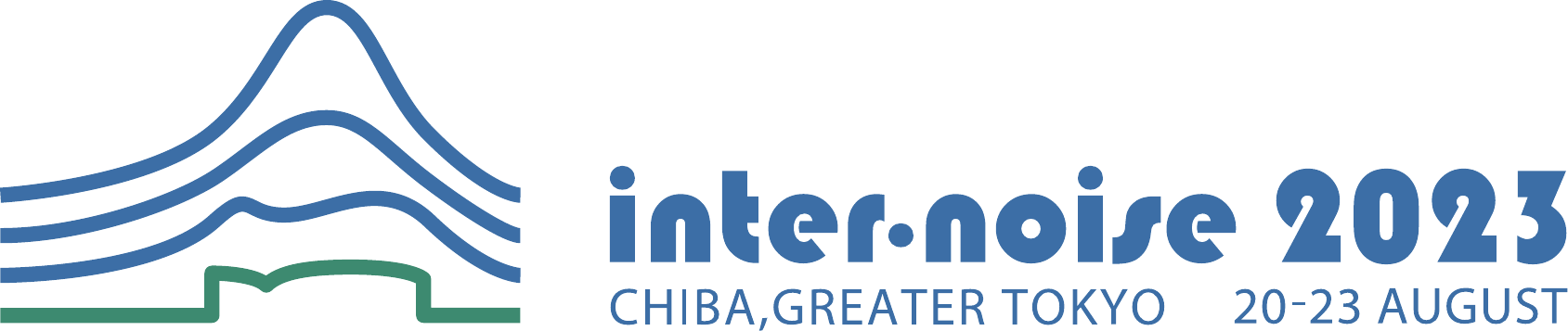}
\end{center}
\vskip.5cm

\begin{flushleft}
\fontsize{16}{20}\selectfont\bfseries
% \textcolor{red}{(The title should be written in "Times New Roman", 16-point, bold font. The first letter of the first word in the title should be capitalized)} \\
Active Noise Control in The New Century: The Role and Prospect of Signal Processing
\end{flushleft}
\vskip1cm

\renewcommand\baselinestretch{1}
\begin{flushleft}

Dongyuan~Shi\footnote{dongyuan.shi@ntu.edu.sg}, Bhan~Lam\footnote{bhanlam@ntu.edu.sg}, 
Woon-Seng~Gan\footnote{ewsgan@ntu.edu.sg}\\
School of Electrical and Electronic Engineering, Nanyang Technological University, Singapore\\
50 Nanyang Avenue, Singapore 639798\\
% Full address\\

\vskip.5cm
Jordan~Cheer\footnote{j.cheer@soton.ac.uk}, Stephen~J.~Elliott\footnote{s.j.elliott@soton.ac.uk}\\
Institute of Sound and Vibration Research, University of Southampton, SO17 1BJ, United Kingdom\\

\end{flushleft}
\textbf{\centerline{ABSTRACT}}\\
\textit{Since Paul Leug's 1933 patent application for a system for the active control of sound, the field of active noise control (ANC) has not flourished until the advent of digital signal processors forty years ago. Early theoretical advancements in digital signal processing and processors laid the groundwork for the phenomenal growth of the field, particularly over the past quarter-century. The widespread commercial success of ANC in aircraft cabins, automobile cabins, and headsets demonstrates the immeasurable public health and economic benefits of ANC. This article continues where Elliott and Nelson's 1993 Signal Processing Magazine article~\cite{elliott1993active} and Elliott's 1997 50\textsuperscript{th} anniversary commentary~\cite{kahrs1997past} on ANC left off, tracing the technical developments and applications in ANC spurred by the seminal texts of Nelson and Elliott (1991), Kuo and Morgan (1996), Hansen and Snyder (1996), and Elliott (2001) since the turn of the century. This article focuses on technical developments pertaining to real-world implementations, such as improving algorithmic convergence, reducing system latency, and extending control to non-stationary and/or broadband noise, as well as the commercial transition challenges from analog to digital ANC systems. Finally, open issues and the future of ANC in the era of artificial intelligence are discussed.}

\section{History of ANC and Key ANC Applications}
Leug's patent, ``Process of dampening sound oscillations", granted in 1936, is often attributed to the dawn of active noise control (ANC) technology~\cite{lueg1936process}. Leug accurately outlined the acoustic noise suppression potential of a well-known acoustical phenomenon: two sound waves with the same frequency and a specified phase difference, when superimposed, will produce destructive interference.

\begin{figure}
    \centering
    \includegraphics[width=15cm]{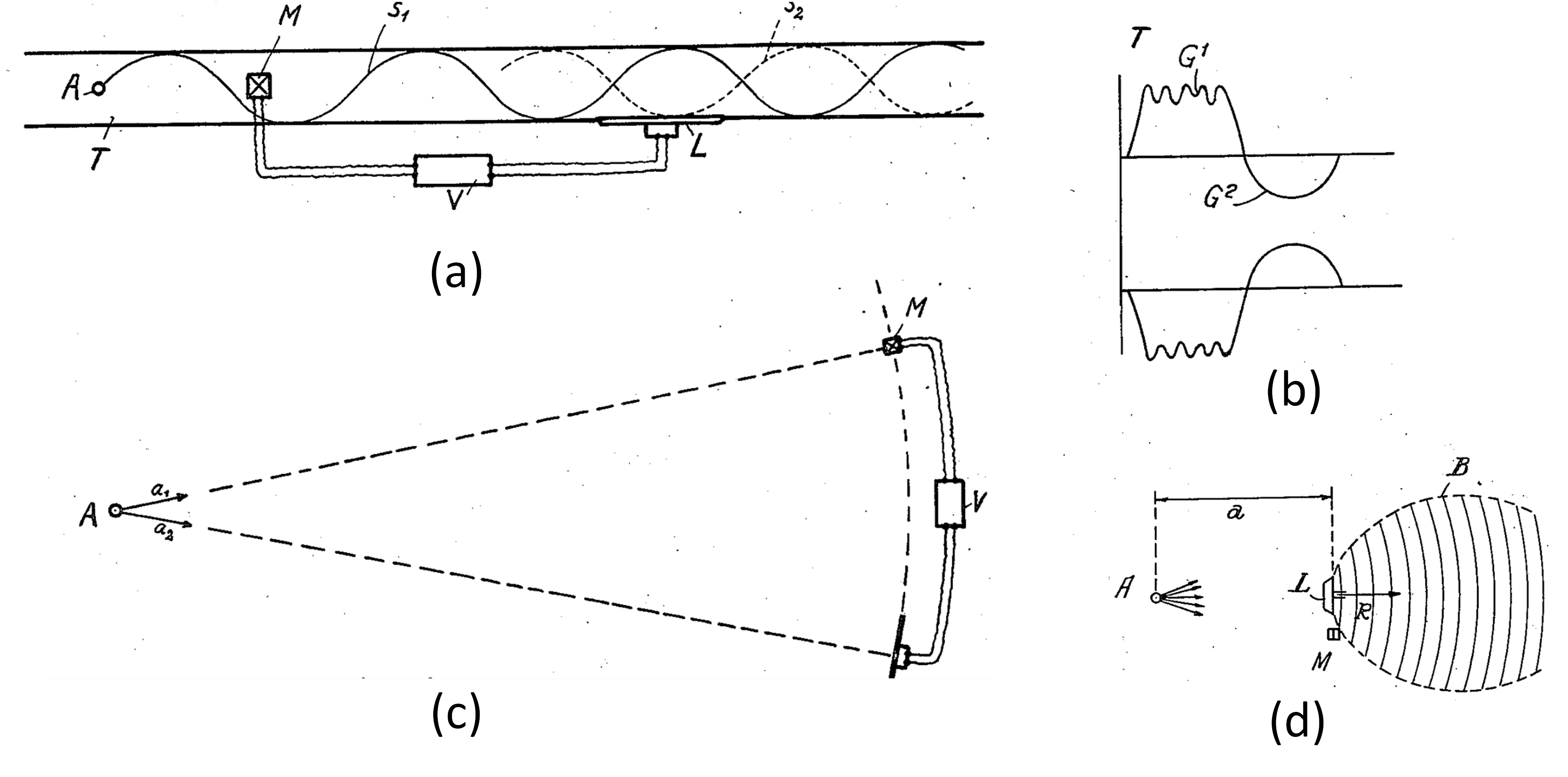}
    \caption{(a) Active noise control duct, (b) the noise and anti-noise wave, (c) the noise source and ANC system, (d) primary noise source and secondary source.}
    \label{fig_patent}
\end{figure}

Owing to the physical nature of the acoustical phenomena, the principles of ANC described by Leug still form the backbone of ANC technology today. Figure~\ref{fig_patent} depicts Leug's now classical formulation of ANC in an air duct $T$, where the microphone/sensor $M$ detects the noise $S1$ produced by the ``primary" noise source $A$. The electronic circuit $V$ processes the sampled noise signal (``reference" signal) and drives the ``secondary" loudspeaker $L$ to generate the ``anti-noise" $S2$. This anti-noise wave $S2$ has the same amplitude as the noise $S2$, but is ${180}^{o}$ out-of-phase, which effectively suppresses the noise. In order to generate the anti-noise $S1$ in time to achieve attenuation, the acoustic delay from $S1$ to $L$ must be precisely measured, and that $M$, $V$ and $L$ must also possess good amplitude and frequency fidelity. Unfortunately, these requirements, which appear simple today, were difficult to meet with the electronics available in the 1930s.
%--------------------------------------------------------------------------------------------------
%--------------------------------------------------------------------------------------------------

Following this pause in advancement, the 1950s was landmarked by two seminal works~\cite{elliott1993active}. Olson and May described an electronic sound absorber consisting of a chamber, a microphone, an amplifier, and a loudspeaker~\cite{olson1953electronic}. This is a ``feedback" variant of Leug's system that reduces the sound pressure level (SPL) due to the noise being near zero to create a ``quiet zone" around the microphone next to the secondary loudspeaker. Around the same time, Conover attempted to reduce the noise produced by a 15000 V transformer~\cite{conover1956fighting} with ANC through a proposed automated tuning process. A 10 dB reduction in SPL on the façade of the transformer was demonstrated but increased the SPL in other areas. Although he was forced to abandon this solution, it inspired others to implement his automatically tuned ANC with analog circuits. However, the inflexibility of analog circuits and the complexity of the debugging procedures drove the field back into obscurity.

The resurgence of ANC was marked by the introduction of microelectronics and landmark algorithmic developments in the late 1970s to 1980s, which sought to overcome the limitations of analog circuitry through adaptive systems. The modern adaptive ANC technique is generally regarded to originate from the adaptive noise canceller proposed by Widrow in 1975~\cite{widrow1975adaptive}. Meanwhile, Morgan and Burgess developed the filter-x least mean square (FxLMS) algorithm \cite{morgan2013history}, which accounts for the acoustic latency in the plant response between the secondary source(s) and the error sensor(s), thereby significantly resolving the convergence issues of the LMS-based algorithm used in ANC applications. Burgess was the first to apply this algorithm to ANC and conducted numerical simulations for noise cancellation in an air duct~\cite{burgess1981active}.

With the invention of the first digital signal processor (DSP) by Texas Instruments in 1978, and the analog-to-digital (ADC) and digital-to-analog (DAC) chips by Intel in 1979, the electronic circuit was transformed from analog to digital. This technological advancement paved the way for the implementation of the adaptive ANC system. For instance, in 1996, Kuo implemented the FxLMS algorithm for ANC on a DSP device that was deployed in a one-dimensional air duct~\cite{kuo1996design}. It obtained approximately $30$ dB of reduction for tonal noise and demonstrated the efficacy of the FxLMS algorithm. The system implementation procedure outlined in the seminal text by Kuo and Morgan~\cite{kuo1996active,kuo1999active} served as a useful starting point for subsequent ANC systems. However, commercial implementations of ANC in ventilation ducts and power transformers eventually faltered in the 1990s due to unsustainable costs in installation, maintenance, and robustness of ANC systems.

Through the mid-1980s to the mid-1990s, ANC research continued to advanced significantly, with the most notable research spearheaded by the Institute of Sound and Vibration Research, University of Southampton in the United Kingdom (Nelson and Elliott, 1993). The research on controlling low-frequency blade-pass noise in aircraft fuselages was successfully demonstrated on the BAe HS 748 propeller aircraft with a 7--13 dB reduction in the noise at the frequencies corresponding to the fundamental and harmonics of the blade passing frequency throughout the cabin~\cite{elliot1990flight}. From this solid foundation, ANC soon found its way into numerous civil and military propeller aircraft~\cite{van1993aircraft,johansson1998active,johansson1999evaluation,johansson2001active, gorman2004active}, most of which are still in service today (e.g. Dornier 328, ATR42/47, A400M, Q400, C-130 variants, and OV-10 Bronco), which inadvertently demonstrated the robustness and stability of ANC in harsh, real-world conditions, as well as the first volume production and implementation of ANC technology. Its success was also perhaps financially and judicially motivated, whereby ANC was necessary to attain safe listening levels without adding too much weight (fuel efficiency) to the aircraft.

During a similar time period, reduction of noise in the automobile cabin using ANC was also investigated. In 1988, Elliott et al applied ANC technology to reduce engine noise~\cite{elliott1988active} and this was subsequently extended to the more challenging road noise control control problem in 1994~\cite{sutton1994active}. Despite Nissan introducing an elegant ANC solution for decreasing vehicle rumble in its 1992 Bluebird in Japan ~\cite{elliott2008review}, the limited attenuation performance and requirement of a separate electronic system meant it did not make economic sense for widespread industry adoption. It was only almost 20 years since the Bluebird when ANC was fully-integrated into the automobile audio system that it found its way back into mass-market luxury vehicles to actively control engine noise in the entire cabin~\cite{hansen2012control}. Despite the success of automobile ANC systems for engine noise, the challenge of controlling the broadband noise associated with road-tyre interactions, eluded its commercial viability for another 10 years. The first mass-produced road noise ANC system, spurred by demand arising from silent engines of electric vehicles, was introduced in 2018 by Hyundai Motors~\cite{oh2018development}. 

Throughout the evolution of ANC, active headsets are arguably the most commercially successful. Intriguingly, work on active headsets also begun in the 1950s with Fogel's patent. This was followed by a long period of inactivity until the introduction of the first commercial ANC headset by Bose in 1989, after 10 years of development. Simultaneously, Sennheiser created their first ANC headset for airline pilots in 1987, the LHM 45 NoiseGard, which demonstrated the viability of analog ANC technology. Throughout the 1990s to 2000s, all active noise-canceling headphones were controlled by analog circuits in a ``feedback" configuration. Analog circuitry is generally regarded to be delayless, is inexpensive, and is simple to manufacture; but its poor accuracy and tunability have hindered its market acceptance. It was only in 2008 that SONY introduced the first digital active noise-cancelling headphone, the MDR-NC500D, which incorporated a digital equalizer to significantly enhance the sound quality of headphones. Since then, digital ANC technology has gradually replaced traditional analog ANC headphones due to its high accuracy, reliability, and configurability to accommodate complex signal processing algorithms.

\begin{figure}[!t]
    \centering
    \includegraphics[width=0.95\linewidth]{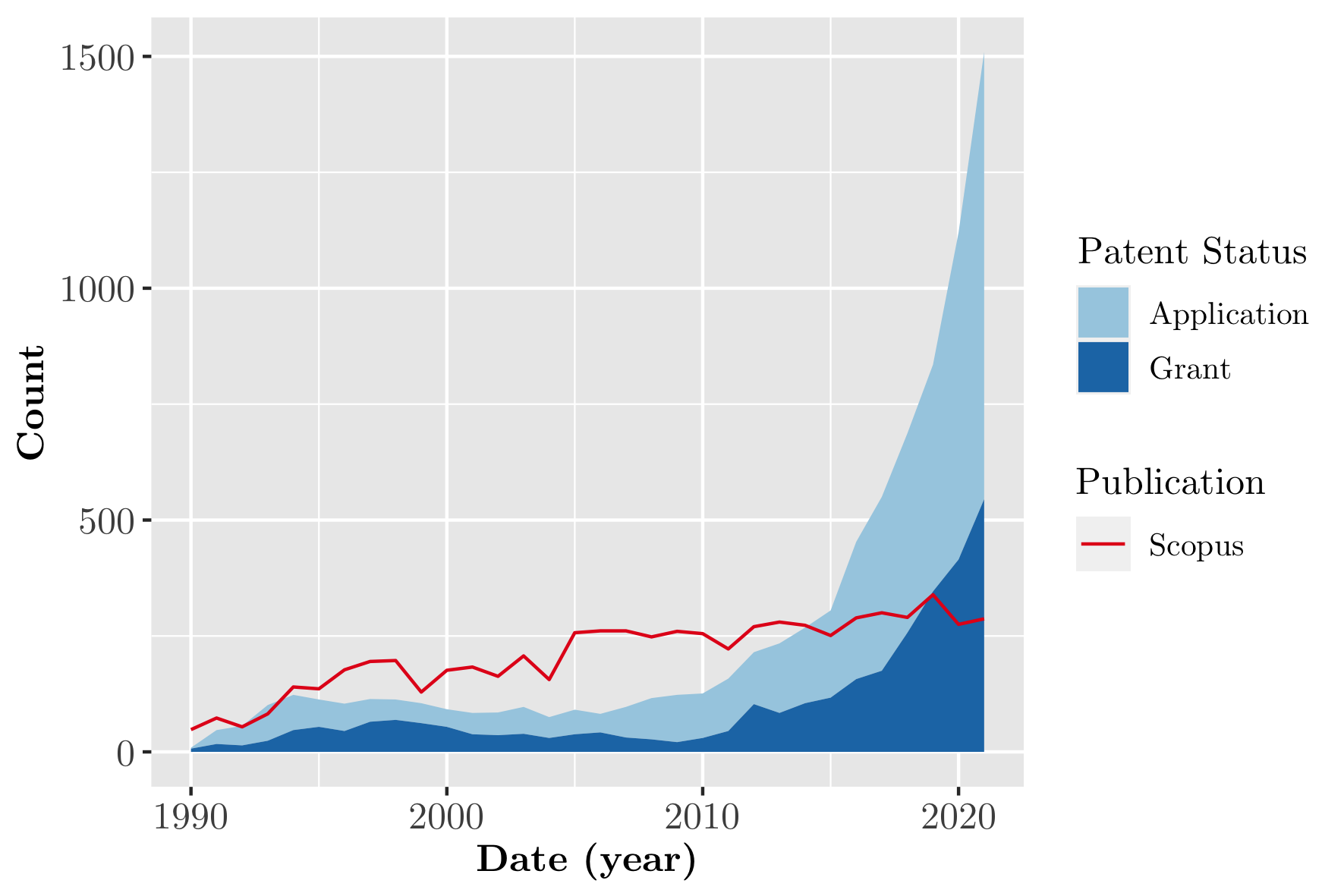}
    \caption{The number of articles indexed in Scopus (\textcolor{myred}{--}); and indexed patent applications (\textcolor{myblue1}{$\blacksquare$}) and granted patents (\textcolor{myblue2}{$\blacksquare$}) from Google patents database  on active noise control from 1990 through 2021.}
    \label{fig_num_articiales_patents}
\end{figure}

As exemplified by the gap in commercial activity in headsets and automobiles, the promise of the 1990s was met with numerous challenges, some of which still remain open problems. Signal processing plays an ever critical role in digital ANC systems of today, and when combined with the knowledge of physical acoustics, electronics, material and mechanical sciences, can help edge the ANC system towards its physical limits. Notably, the boom in academic ANC research has persisted since the 1990s, but it was only in the 2010s that we experienced an exponential commercial interest. The trends in publications and patent activity are based on keyword searches in Scopus and Google patents databases, respectively, as illustrated in Figure~\ref{fig_num_articiales_patents}.
%--------------------------------------------------------------------------------------------------
%--------------------------------------------------------------------------------------------------

\section{Scientific Challenges  and Milestones}
Digital electronic systems are gradually replacing traditional
analog circuits because of the booming semiconductor industry, and particularly the invention of DSP processors and ADC/DAC converters~\cite{shi2016comparison}. Digital ANC systems have become commonplace in academia and industry owing to their high degree of tuneability and adaptability, and the relative ease of design. This advancement lays the groundwork for the implementation of more complex and efficient control algorithms for ANC systems, providing potential such as improved performance, lower cost, and a greater variety of applicability and functionality. The proposal and development of adaptive algorithms, such as the FxLMS algorithm, made it possible for ANC systems to handle time-varying systems and noise sources.
However, there are still a number of scientific and practical obstacles that impede its development. In practice, the ANC system places a high processing demand on the real-time processor, which must complete the reference signal sampling, processing, and generation of the anti-noise rapidly so that the anti-noise can effectively interact with the acoustic noise to attenuate it. This requirement poses a causality constraint on practical ANC systems. The challenge associated with the causality constraint is exacerbated in portable ANC systems, such as ANC headphones. Their diminutive size reduces the acoustic delay between the reference microphone and the error microphone, thereby limiting the available processing time window. Due to limitations of cost, size, and power, it is difficult to install a powerful processor in ANC headphones. This is the primary reason why current commercial headphones typically filter with fixed coefficients during ANC, despite the fact that adaptive algorithms offer significantly superior noise reduction performance~\cite{shen2022adaptive,shen2022multi}.

For the portable devices~\cite{chang2016listening}, the integration of ANC with other audio and communication functions, such as transparent listening, occlusion removal, acoustic echo cancellation, and focused listening, among others, would open the door to a variety of intriguing applications in hearables. This integration brought up many interesting challenges, including how to automatically activate certain functions in response to environmental noise. 

Another challenge associated with ANC systems is that associated with nonlinearity~\cite{shi2017effect,shi2021optimal,shen2023momentum,shi2019two,shi2019optimal,shi2023multichannel}. Nonlinearities are common in practical ANC systems due to imperfections, such as limited power capabilities of transducers used for sensing and actuation, and both vibration and acoustic propagation paths. Such nonlinearities can lead to divergence in conventional adaptive algorithms~\cite{shi2017effect}. Therefore, many nonlinear adaptive algorithms have been proposed to address this issue~\cite{lu2021survey}, but due to their high computational complexity, they have usually prevented them from being used in practical scenarios.

To reduce the computational load on the real-time processor, many computationally-efficient adaptive algorithms~\cite{shi2021block,shi2019practical}, have been proposed for use in mid-size ANC systems, which are typically used to achieve global noise reduction throughout a space, such as a bedroom, an airplane cabin, or a car's interior. Even though some modified adaptive algorithms can achieve a balance between convergence and computation cost, their less-than-ideal response time to rapidly varying noise still diminishes their perceptual instantaneity of noise reduction.

Perhaps the most technically challenging, but arguably the most societally impactful is to achieve a large quiet zone with multichannel ANC (MCANC) systems in a fully- or partially-open space. Multiple-input, multiple-output (MIMO) technology and the corresponding adaptive algorithms have been incorporated into MCANC systems. There are, however, significant differences between MCANC and standard MIMO systems, such as those used in wireless communication. Typically, the performance of the MCANC system is limited by the boundary conditions of the complex acoustic environment. Traditional MCANC algorithms are primarily concerned only with the time-domain signal acquired by the microphones but are not conditioned to consider the acoustic boundary conditions of the whole control region. This diminishes the global control efficacy of the MCANC system in a large control region, even potentially resulting in a ``spillover", increasing the sound pressure outside the control region. Despite the fact that recent spatial ANC techniques transform the sound wave from the time-domain to the wave-domain~\cite{zhang2018active}, thereby incorporating acoustic boundary information and providing more efficient control over the space, their high computational requirements limit their use in real-time applications.

In addition, the positions of the microphones and secondary sources are crucial to MCANC system performance in realistic scenarios. The conventional trial-and-error method is ineffective when applied to large-scale ANC systems with multiple control units. Some solutions based on metaheuristic optimization procedures, such as the genetic algorithm (GA), still require a substantial amount of human intervention in the measurement of the sound field.

It is worth noting that in traditional ANC research, humans are often disregarded in the design of the controller, which simply focuses on minimizing the mean square of the error signal(s), for example. Therefore, an increasing number of researchers are examining this topic and attempting to combine psycho-acoustic concepts with conventional active control methods, such as noise masking and soundscape techniques~\cite{lam2021ten}. Nonetheless, these techniques are still in their infancy and have not been broadly validated and implemented.

\section{Impacts in the society}
With cities rapidly urbanizing, our living space is becoming increasingly congested with intricate transportation networks, constant construction activities, and busy industrial zones, all of which generate significant and unwanted noise. Following this, urban acoustic noise ceaselessly afflicts humans in their everyday lives and thus impacts the health of the population. Low-frequency noise can have a significant impact on public health since it is not easy to control using traditional passive noise control solutions. However, ANC is most effective at reducing low-frequency noise, and can thus have a significant impact on public health. ANC systems can be implemented within a small form factor, and are thus widely used in portable audio devices, such as headphones, which can be used in any environment and thus have a significant influence on reducing the impact of environmental noise on the population. ANC has also gradually become a technology of significant interest within industrial and commercial applications, thus generating both significant research challenges and a massive market requirement.
The core component of ANC technology is the algorithms, which have been developed, and continue to evolve, through decades of academic research in the field of signal processing. Although there are several successful ANC solutions in the market, present ANC technology, which is based on conventional signal processing techniques, still faces numerous practical obstacles. To address these practical problems, there have been many novel algorithms and control strategies proposed recently and these contributions lie directly within the academic research fields linked to the SPS community and its publications, ANC is also an important field. Meanwhile, there are many novel algorithms and control strategies are proposed recently to address these practical problems. Furthermore, due to the intricacy in the application of ANC technologies, it has also become a focal point for interdisciplinary research and in recent years has brought signal processing research together with acoustics, vibration, control, psycho-acoustics, machine learning, human-machine interaction, sensing, actuation, and many more areas.

\section{Current perspectives and recent advancements in ANC}
With the rapid advancement of artificial intelligence and its proliferation throughout SPS domains, several researchers have attempted to use deep learning algorithms to realize ANC systems that overcome practical challenges or offer augmented capabilities. To increase the efficacy of noise reduction for dynamic noise and reduce the impacts of system nonlinearity, complex neural networks have been utilized to directly replace the control filter and process the reference signal~\cite{zhang2022attentive, zhang2021deep, zhang2023deep,zhang2023low,luo2022implementation}. However, the limited computing capability of real-time controllers confines such computationally intensive deep-learning techniques to simulations. 
% \sout{A new approach based on the selective fixed-filter ANC (SFANC) technique uses convolutional neural networks (CNN), but avoids this computational complexity issue. The SFANC technique employs a lightweight CNN model running on a coprocessor to assist the real-time controller in selecting the most appropriate pre-trained control filter from a library based on different types of noise. As a result, this method can reduce the controller's computational load  compared to an fully adaptive controller and increase the system's stability, as it does not involve any feedback process.} 
The computational complexity issue can be circumvented by a novel method that employs a lightweight convolutional neural network (CNN) model to select a pre-trained control filter based on the noise type. As the CNN model runs asynchronously on a co-processor using block computation and the real-time controller only executes the normal filter, a significant overhead of real-time computations is effectively avoided. In addition, since there is no feedback process involved, this method also increases the system's stability~\cite{shi2022selective, shi2020feedforward,luo2022hybrid,shi2023transferable}. Notably, current manufacturers incorporate a filter-selection mechanism into their ANC chips, paving the path for the deployment of the deep-learning-based SFANC approach~\cite{luo2023deep,luo2022performance}. Furthermore, the conventional adaptive algorithm's hyper-parameter selection, such as step size and initial filter settings, affects the ANC system's convergence and stability. Data-driven approaches, such as Meta-learning~\cite{shi2021fast}, have also been employed to learn these parameters for a noise dataset automatically. This strategy is a good solution to the conventional approach to determining the hyper-parameters through trial and error-based tuning.

Recent spatial ANC technology utilizes the spherical harmonic-based decomposition to move the sound signal from the time domain to the wave domain in order to improve the noise control zone in the free field~\cite{zhang2018active,koyama2021spatial}. Utilizing the spatial information of the sound field, it is possible to create a wide zone of noise reduction. Typically, this strategy is utilized when the confined zone has a consistent shape, such as a spherical or columnar region. An increasing number of researchers are attempting to broaden the application of this approach and to develop effective computing methods for its implementation.

When attempting to control noise actively within a zone for a particular user, such as the driver in an automobile cabin, the size of the generated quiet zone decreases with increasing frequency. This effectively limits the upper frequency of control, particularly when the user moves. To overcome this problem, head-tracking technology, based on image processing, has recently been used along with remote sensing strategies~\cite{zhang2021robust,shi2019selective,lai2023real,shi2020feedforward_virtual} to move the position of the zone of quiet with the user and thus increase the control bandwidth~\cite{elliott2018head}. This technology has been applied within the automotive environment~\cite{jung2019local}, but has many more potential applications.

There have been several other unique ANC applications in recent years~\cite{kajikawa2012recent}, including the ANC window system, which employs secondary sources implanted in the window frame to reduce incoming noise while maintaining natural ventilation and light ingress~\cite{lam2020active,hasegawa2018window}. ANC techniques have also been used in conjunction with sound barriers to improve traffic noise cancellation performance~\cite{qiu2019introduction}. ANC systems have also been utilized to cancel the high sound levels of the machinery, including the diverse applications of construction machines~\cite{wen2021design,shi2022integration}, home appliances~\cite{mazur2019active,lee2021review}, and the fMRI scanner~\cite{reddy2007analysis}, with the aim of creating a quiet working environment for the operator or user. Some other applications attempt to eliminate acoustic noise in confined spaces, with novel systems incorporating an ANC system into a pillow to reduce snoring sounds~\cite{chang2013active}. Without the physical restrictions of wires, wireless sensors are also used in the ANC system and placed closer to the noise sources in order to gather the reference or error signal with a higher signal-to-noise ratio~\cite{shen2023implementations,shen2021wireless,shen2022hybrid}. Many distributed ANC strategies have been created that take advantage of the cutting-edge decentralized technique to assign the computations to each sub-computing node in order to build the large-scale ANC system while reducing the huge computational complexity~\cite{ferrer2023assessment,ji2023practical}.

\section{Conclusions}
The proposed feature paper aims to provide a systematic review of the evolution of ANC technology over the past quarter-century via the lens of signal processing. The application of signal processing research results to the ANC sector is demonstrated to the reader. This article summarizes the main application areas and academic research results of ANC technology till now.  It outlines the technical bottlenecks and opportunities encountered and looks forward to its future developments.

\section*{Acknowledgements}
\noindent This research/work is supported by the Singapore Ministry of National Development and National Research Foundation under the Cities of Tomorrow R$\&$D Program: COT-V4-2019-1. Any opinions, findings, and conclusions or recommendations expressed in this material are those of the author(s) and do not reflect the views of the Singapore Ministry of National Development and National Research Foundation, Prime Minister’s Office, Singapore.

\bibliographystyle{unsrt}
\bibliography{sample} 

\end{document}